\documentclass[12pt]{iopart}
\usepackage{iopams}  
\expandafter\let\csname equation*\endcsname\relax
\expandafter\let\csname endequation*\endcsname\relax

\usepackage{cite}
\usepackage{graphicx} 
\usepackage{dcolumn}
\usepackage{bm}
\usepackage{mathtools}
\usepackage{hyperref}
\usepackage{graphicx}
\usepackage{amsmath}
\usepackage{braket}
\usepackage{amsfonts}
\usepackage{amssymb}
\usepackage{bm}
\usepackage[usenames,dvipsnames]{color}
\usepackage{float}
\usepackage{amsmath,amssymb}
\usepackage{amsfonts}
\usepackage{csquotes}
\usepackage{comment}
\usepackage{bbold}
\usepackage{romannum}
\usepackage{mathrsfs}
\hypersetup{
	colorlinks=true,
	citecolor=blue,
	linkcolor=blue,
	urlcolor=blue}

\newcommand{\onlinecite}[1]{\cite{#1}}

\usepackage{subfig}
\usepackage{tikz}
\usepackage{pgffor} 
\usepackage{hyperref}

\begin{document}

\title[]{Current-induced spin polarisation in Rashba-Dresselhaus systems under different point groups}

\author{Akash Dey$^1$$^,$$^2$$^,$$^*$, Ashis K. Nandy$^1$$^,$$^2$, Kush Saha$^1$$^,$$^2$}

\address{$^1$ National Institute of Science Education and Research, Jatni, Odisha 752050, India}
\address{$^2$ Homi Bhabha National Institute, Training School Complex, Anushakti Nagar, Mumbai 400094, India}
\address{$^*$ Author to whom any correspondence should be addressed.}

\eads{\mailto{akash.dey@niser.ac.in},   \mailto{aknandy@niser.ac.in},
\mailto{kush.saha@niser.ac.in}}

\date{\today}

\begin{abstract}
Non-magnetic materials without inversion symmetry typically exhibit strong Rashba spin-orbit coupling (SOC), enabling the well-known Rashba Edelstein effect where an external electrical current induces transverse spin polarisation. In this study, we demonstrate that electrically induced spin polarisation in non-magnetic materials, for example, electronic systems within quantum-well geometries, can significantly be influenced by the system's point-group symmetries, such as $C_n$ and $C_{nv}$. These symmetries allow various linear and higher-order momentum, $k-$varying SOC Hamiltonian.
Specifically, we show that surfaces having $C_{n}$ point-group symmetry, which permits specific linear and cubic Rashba and Dresselhaus SOC terms, can lead to both orthogonal and non-orthogonal spin polarisations with respect to the applied field. In contrast, surfaces with $C_{nv}$ symmetry exhibit only transverse spin polarisation, regardless of the linear and cubic SOC  terms. We further find contrasting spin polarisation for cubic-in-$k$ SOC as compared to the linear-in-$k$ SOC when energy is varied, for example, through doping. Additionally, we show that the surfaces with $C_{n}$ symmetry may exhibit persistent spin current, depending on the relative strength between different momentum-dependent SOC terms. Our finding emphasizes the significance of crystal symmetry in understanding and manipulating induced spin polarisation in noncentrosymmetric materials, especially in surface/interface systems. 
\end{abstract}

	
\maketitle
	
\section{\label{sec:level1}Introduction}
Creating dissipationless transport is a central goal in modern quantum transport phenomena, particularly in generating and tunneling of pure spins using electrical currents within quantum devices~\cite{vzutic2004spintronics,hirohata2020review,wu2010spin}. The key quantity that controls such spin generation and/or transport in materials with broken inversion symmetry is the spin-orbit coupling (SOC)~\cite{bychkov1984properties}. This SOC term that linearly depends on momentum in the presence of structure-inversion asymmetry (SIA) gives rise to an isotropic splitting of electronic bands{---}the linear Rashba effect~\cite{Au111_PRL1996, Au111_PRB2011,bychkov1984oscillatory,rashba1960spin, Bihlmayer_2015}. An analogous band splitting in the presence of bulk-inversion asymmetry (BIA) also occurs due to the Dresselhaus spin-orbit coupling (DSOC)~\cite{DresselhausPR1955}. 
The utilization of such SOC terms in non-centrosymmetric materials has led to numerous nontrivial quantum states, transport phenomena, and interaction effects, promising for potential applications in devices~\cite{AKN_NatComm2016,soumyanarayanan2016emergent, AKN_PRL2016, DMI-PRB2023}. The Rashba Edelstein effect (REE) is one of such phenomena where a non-equilibrium spin polarisation is induced by an electrical current due to the presence of Rashba spin-orbit coupling (RSOC)~\cite{edelstein1990spin,vignale2016theory,silsbee2003erratum,aronov1989nuclear, Perkins_2024, Johansson_2024}. Since its first prediction, a great volume of work has focused on studying the REE effect in semiconductors, metallic surfaces and later extending it to the topological materials~\cite{kato2004current,silov2004current,zhang2015charge,zhao2020unconventional,zhao2020observation,lv2018m,stern2006current,Chen_2020}. Similarly, a related phenomenon, namely orbital Edelstein effect as predicted in bulk antiferromagnets~\cite{AKN_NatComm2019} has attracted considerable interest in recent time. Nonetheless, the REE can also lead to spin-orbit torques, which emerge as a promising candidate for the electrical manipulation of magnetic order~\cite{Wadley_Science2016,gambardella2011current}.

In non-centrosymmetric systems, the RSOC for degenerate atomic orbitals such as $p$, $d$, and $f$ orbitals induces a momentum-dependent spin splitting. This splitting can extend beyond the linear regime, giving rise to more complex physical phenomena. Often, these systems exhibit a cubic dependence of band splitting on momentum ($k$)~\cite{vajna2012higher,nakamura2012experimental, AnomalousRashba_PRB2002,moriya2014, Schulz_PRB2021, Maleki}. Unlike the linear-in-$k$ SOC that induces uniform splitting in Rashba bands, the cubic counterpart creates a directionally dependent energy separation that varies with momentum. Moreover, due to the presence of specific cubic terms allowed by the crystalline symmetry, materials with $C_{3v}$ or $C_{4v}$ point group symmetry exhibit anisotropic spin splitting~\cite{koroteev2004strong,sugawara2006fermi,ast2007giant,moreschini2009assessing,gierz2010structural,mirhosseini2010toward}.  Various electron and hole systems have been experimentally found to exhibit spin-splitting anisotropic in $k$~\cite{winklerbook,gerchikov1992,marcellina2017}. For example, SrTiO$_3$ surface~\cite{nakamura2012experimental}, SrTiO$_3$-based asymmetric oxide heterostructures~\cite{lin2019}, Ge/SiGe quantum well~\cite{moriya2014}, antiferromagnet TbRh$_2$Si$_2$~\cite{usachov2020} are known to show non-linear spin splitting in their surface/interface electronic structures.
Additionally, recent research in bulk insulators has revealed materials exhibiting purely nonlinear RSOC where various spin textures (STs) govern controlled manipulation of dissipationless transport. In Ref.~\onlinecite{johansson2016theoretical}, current-induced spin polarisation has been studied in $C_{2v}$ surface with only mass anisotropy and in $C_{3v}$ surfaces with out-of-plane cubic term in addition to the in-plane RSOC. Similar cubic RSOC has been used in Ref.~\onlinecite{liu2008current} and in Ref.~\onlinecite{mawrie2014} to discuss spin polarisation and conductivity. 
Despite the intriguing physics observed in higher-order momentum terms under various point group symmetries, a comprehensive understanding of their influence on induced spin in the presence of electric fields remains elusive. Notably, a systematic investigation is lacking that explores this effect across both symmetry-preserved and symmetry-broken materials. This is expected to be of relevance for the interpretation of possible spin-polarisation measurements detecting the Rashba spin splitting beyond the linear limit.

In view of the above, we provide a detailed and comprehensive study of current-induced spin polarisation in a two-dimensional electron gas (2DEG) under different point group symmetries, particularly $C_{3/{3v}}$ and $C_{4/{4v}}$ symmetries. We note that semiconductor quantum well structures and two-dimensional (2D) nanostructures with carefully chosen growth geometries can host 2DEG with such symmetries.
By employing the Kubo formalism, we show that under $C_3$ or $C_4$ symmetry, a combination of standard RSOC and a symmetry-allowed linear term (dubbed as DSOC($+$)) leads to isotropic spin polarisation, independent of their relative strengths.
This is in contrast to the case where both Rashba and standard DSOC, namely DSOC$(-)$, together lead to anisotropic spin polarisations~\cite{maxim2007,schliemann2003nonballistic, Bernevig, SchliemannJohn, Tao}. 
We further supplement these findings by solving the time-dependent Schr\"odinger equation, particularly showcasing the evolution of spin polarisation under the influence of an in-plane electric field applied in any direction. Remarkably, we find that the components of spin-polarisation switches sign along the orthogonal directions of the applied field for DSOC($+$) in contrast to the DSOC($-$) where no such sign change is found (Sec.~\ref{subRSOCDSCOlin}). Moreover, we find a special situation where the collinear alignment of Rashba and DSOC($-$) leads to the persistent spin texture (PST). This, in turn, gives rise to zero spin polarisation in the presence of the applied field. 

Similar to the linear term, purely cubic RSOC and DSOC terms together under $C_3$ or $C_4$ point group symmetry can host both orthogonal and non-orthogonal spin polarisation in the presence of an electric field. However, the magnitude of the spin polarisation increases with the Fermi energy as opposed to the linear case where it reduces with the Fermi energy due to the partial compensation of spin polarisation from carriers of opposite chirality (Sec.~\ref{subRSOCDSOCnonlin}). 
Interestingly, in symmetry-enriched electronic systems characterized by $C_{3v}$ or $C_{4v}$ symmetries, we find only transverse polarisation even in the presence of cubic RSOC. However, the magnitude of spin polarisation under $C_{4v}$ surfaces substantially differ from that of systems with $C_{3v}$ symmetry (Sec.~\ref{subsubc3v}). Specifically, we find that the anisotropy of the Fermi contours (FCs) overall reduces the magnitude of spin polarisation in $C_{4v}$ and it strongly depends on the orientation of the applied field. Additionally, the $C_{4v}$ symmetry allowed mixed cubic RSOC leading to net zero polarisation due to the deformed nature of the FCs (Sec.~\ref{subsubc4v}), regardless of the direction of the applied field. This effect is also manifested in the total spin polarisation as we vary Fermi energy. 
We finally conclude with a discussion on how the spin polarisation as a function of the direction of the electric field can assist in identifying the symmetry properties of the materials.

\begin{figure}
\centering
\includegraphics[width=0.75\linewidth]{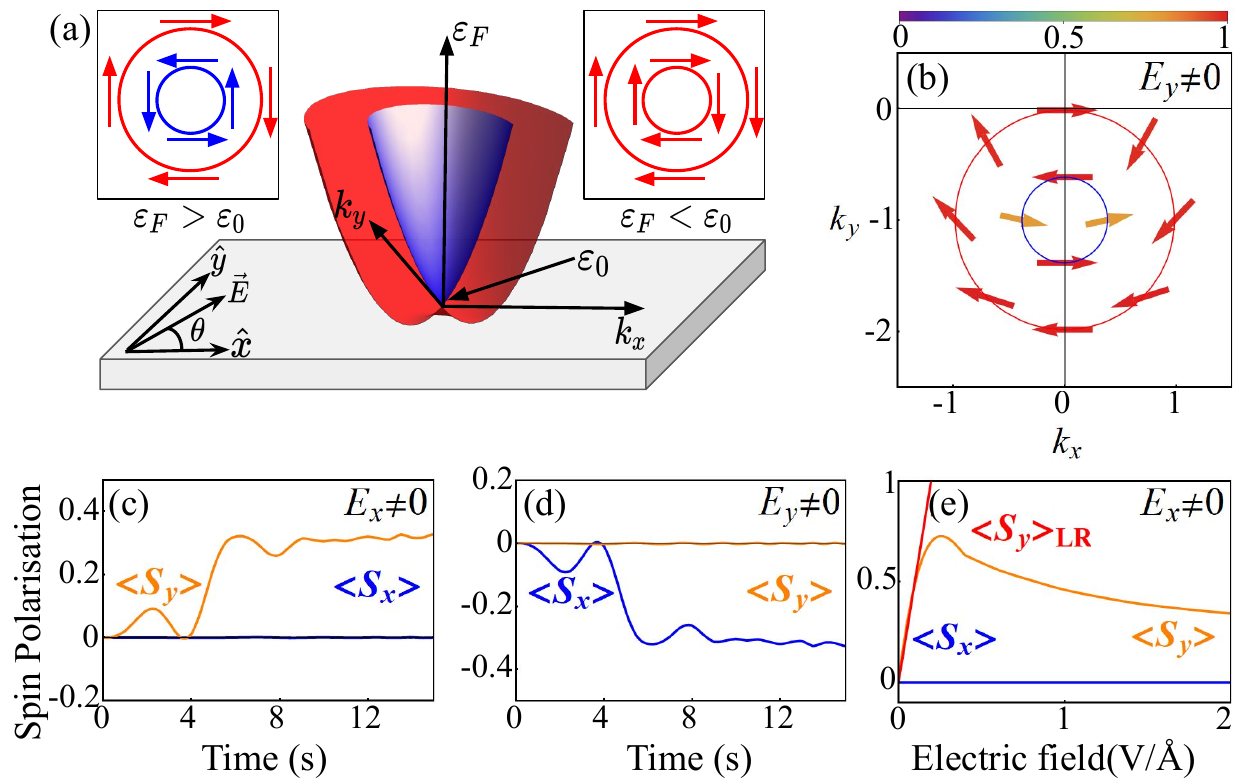}
\caption{(a) The schematic of a 2DEG setup under a static electric field $\textbf{E}$. With the linear RSOC, a free electron state splits into two subbands with opposite spin textures (red and blue) that intersect at the BCP. The corresponding energy is defined as $\varepsilon_0=0$. The left inset shows spin orientation along FCs for $\varepsilon>\varepsilon_0$, and the right inset shows the same for $\varepsilon<\varepsilon_0$ at zero electric field. (b) Snapshot of spin polarisation at a particular instant of time ($t=4$s) for the electric field along $y$, obtained by solving the time-dependent Schr\"odinger equation. Evidently, the FCs move in the opposite to the applied field. 
r(c-d) Time evolution of momentum-averaged (over both the FCs at $\varepsilon_\textrm{F}= 0.63$ eV) spin polarisation $\langle S_x\rangle$ and $\langle S_y\rangle$ per unit area for the field along $x$ and $y$, respectively. (e) Time and momentum-averaged spin polarisation as a function applied field along $x$. For weak fields, it varies linearly, corroborating the results ($\langle S_y\rangle_\textrm{LR}$) obtained from Kubo formalism (for the outer FC). Here, we take $\alpha=1$ eV.$\textrm{\AA}$. Also, throughout the text, we use $\hbar^2/2m=$1.66 eV.\AA$^2$, $E=$0.25 V/\AA~ and $k_x$ and $k_y$ in units of \AA$^{-1}$; the color bar shows the magnitude of each spin polarisation.}
\label{fig:spin_textureROC}
\label{analytical}
\end{figure}

\section{\label{sec:level2} Theoretical Background}
We begin by reviewing the REE within a 2DEG model system~\cite{rashba1960spin}.
The corresponding Hamiltonian with the linear-in-$k$ RSOC term is 
\begin{align}
H_0=\frac{\hbar^2 \mathbf{k}^2}{2 m}+\alpha(k_x \sigma_y -k_y \sigma_x),
\label{eq:ham0}
\end{align}
where $\mathbf{k}\in (k_x,k_y)$ represents the 2D crystal momentum, $\boldsymbol{\sigma}$ denotes the Pauli matrices, and $\alpha$ corresponds to the linear Rashba coupling parameter. 
This linear coupling induces a spin-splitting in the energy bands that scales linearly with the crystal momentum as shown in Fig.~\ref{fig:spin_textureROC}(a). This is typically termed as the linear Rashba splitting. In standard Rashba Hamiltonian, $\alpha$ is given by the ratio $\alpha = 2E_\textrm{R}/k_\textrm{R}$ between the spin splitting energy $E_\textrm{R}$ and the momentum offset $k_\textrm{R}$~\cite{Sino,Chen}. Insets of Fig.~\ref{fig:spin_textureROC}(a) shows the spin helicities on the inner and outer FCs at zero field at two different energies $\varepsilon>\varepsilon_0$ (left) and $\varepsilon<\varepsilon_0$ (right), where $\varepsilon_0 =0$ refers to the band crossing point at zero momentum. Clearly, the spins are perpendicular to the motion of electrons, indicating the spin-momentum locking of the system. Moreover, for $\varepsilon>\varepsilon_0$, spins in the inner and outer FCs are found to align antiparallel to each other as opposed to the case for $\varepsilon<\varepsilon_0$. We note that the magnitudes of spins in each FC depend on $\varepsilon_\textrm{F}$, which in turn modify the induced $\langle S^{in}\rangle$ and $\langle S^{out}\rangle$ in the presence of electric field as will be discussed shortly.

\begin{figure}
\centering
\includegraphics[width=0.66\linewidth]{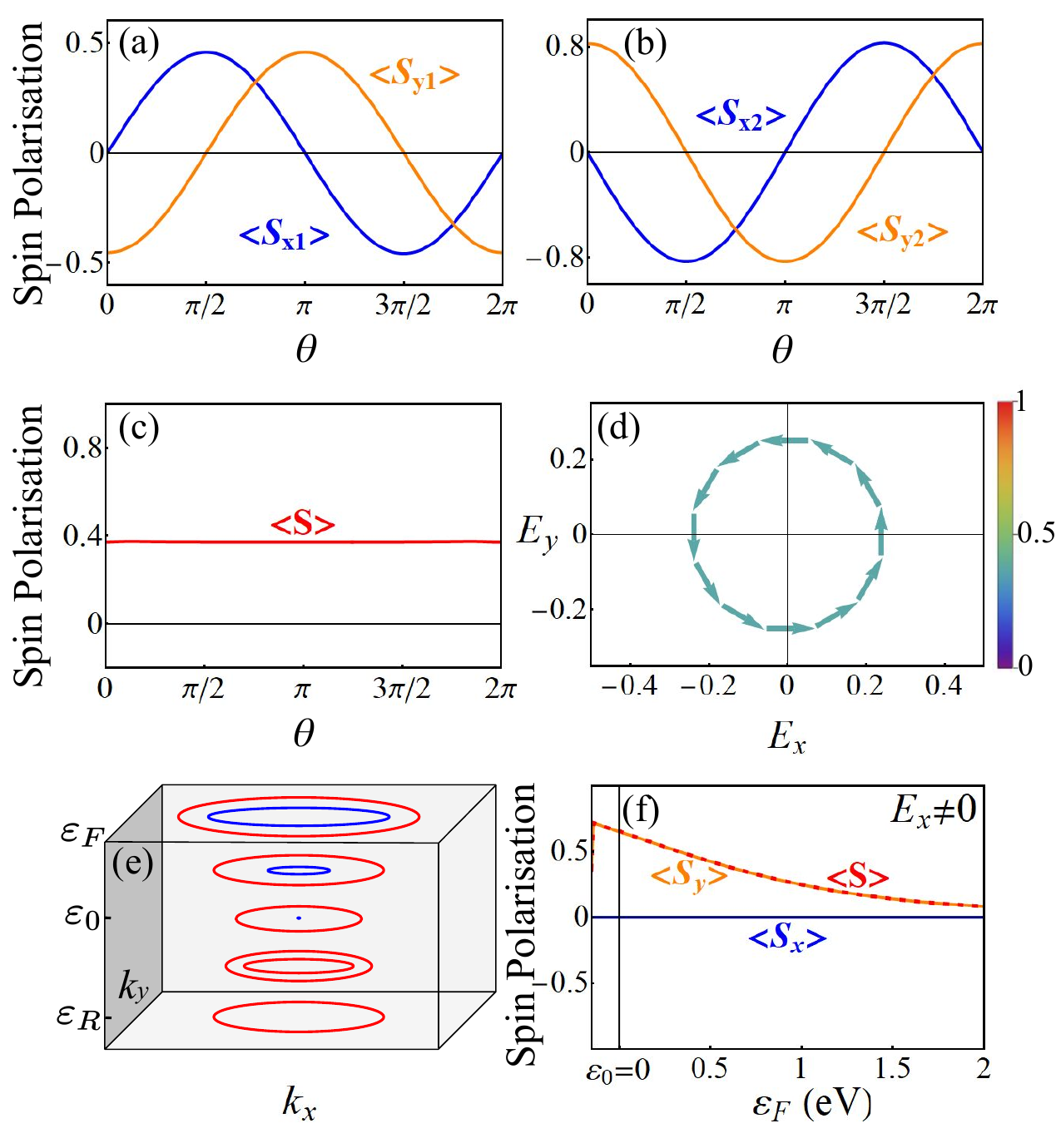}
\caption{(a-b) $x$ and $y$ components of spin polarisation per unit area as a function of the direction of the applied field for inner (1) and outer (2) FCs at $\varepsilon_\textrm{F}=0.63$~eV, (c) Total time-averaged and momentum averaged (over both the FCs at $\varepsilon_\textrm{F}$) spin polarisation as a function of the direction of the applied field, (d) Resultant time and momentum averaged spin polarisation as a function of electric field in the  $E_x$-$E_y$ plane. (f) Fermi energy dependence of different spin components and total spin polarisation for the electric field along $x$. The vertical line represents $\varepsilon_\textrm{F}=\varepsilon_0$. (e) illustrates FCs as a function of energy and the color code refers to the relative orientation of the spin in both the FCs.}
\label{fig:spin_textureElectric}
\end{figure}

Using the Kubo formalism, we first compute the spin polarisation in the presence of an external static electric field $\mathbf{E}$. In the linear response regime, the spin polarisation in the frequency domain can be expressed as~\cite{Bruus2004,mahan2013many, Li},
\begin{align}
\langle S_i(\omega)\rangle=&-\frac{eE_j}{\omega}\int\frac{d\varepsilon}{2\pi}f(\varepsilon)\sum_k {\rm Tr}\left[\hat{v}_jG_A(\varepsilon-\hbar\omega,\mathbf{ k})\hat{S}_i\right.\nonumber\\&
[G_R(\varepsilon,{\bf k})-G_A(\varepsilon,{\bf k})]+\hat{v}_j[G_R(\varepsilon,\mathbf{k})-G_A(\varepsilon,\bf k)]\nonumber\\
&\left.\hat{S_i}G_R(\varepsilon+\hbar\omega,\bf k)\right], 
\end{align}
where $(i,j)\in (x,y,z)$, $e$ is the electron's charge and $f(\varepsilon)$ is the Fermi distribution function. $\hat{v}$ and $\hat{S}$ are the velocity operator and the spin operator, respectively. $G_R$ and $G_A$ are the retarded and advanced Green functions, respectively. $G_{R}^{-1}(\omega,{\bf k})_{nn'}$ ($G_{A}^{-1}(\omega,{\bf k})_{nn'}$) is defined as $(\hbar\omega-H + i\eta)$ ($(\hbar\omega-H - i\eta)$), where the broadening parameter $\eta$ represents the inverse of the relaxation time $\tau$ and takes into account the finite lifetime of electron states with $n,n'$ as their band indices. 
At zero temperature, $i$th of component of the induced spin polarisation can be written as (restricting to the linear expansion in $\omega$ of the Green's function)
\begin{align}
\langle S_i(\omega)\rangle=&\frac{eE_j}{2\pi}\sum_k {\textrm{Tr}}[\hat{v}_jG_A(\omega)\hat{S}_iG_R(\omega)].
\label{eq3}
\end{align}

In case of an applied electric field along $x$, $i.e.$ $\textbf{E} = (E_x,0,0)$, a straightforward calculation (Appendix~\ref{app:kubo1}) leads to
\begin{align}
&\langle S_y\rangle\approx \frac{e E_x \varepsilon_\textrm{F}}{32\pi\alpha \eta}, 
\langle S_x\rangle=\langle S_z \rangle=0,
\label{eq:spin_Rashba}
\end{align}
where $\varepsilon_\textrm{F}$ is the Fermi energy. Thus, we obtain $\textbf{S} = (0, S_y,0)$, indicating a transverse spin polarisation, and we recover the conventional REE. This can also be understood from Eq.~(\ref{eq:ham0}) itself, where one can rewrite the Rashba term as $ H_{\rm R}=\textbf{B}_\textrm{eff}\cdot \boldsymbol{\sigma}$ with $\textbf{B}_\textrm{eff}=\alpha(-k_y,k_x)$. The static electric field along $\hat{x}$ couples to the $y$ component of $\textbf{B}_\textrm{eff}$ via Peierls substitution $k_x\rightarrow k_x+A_x$, $A_x$ being the $x$-component of the gauge potential due to the field. This, in turn, leads to the spin polarisation along $\hat{y}$.

This outcome can be further validated by studying the temporal evolution of the induced spin polarisation under the applied electric field $\textbf{E}(t)$. This involves solving the time-dependent Schr\"odinger equation $i\partial_t H|\psi\rangle=\varepsilon|\psi\rangle$ with the gauge potential $\partial_tA(t)=-e \textbf{E}(t)\cdot\hat{\boldsymbol{r}}$, where $\hat{\boldsymbol{r}}$ is the position operator. Notably, the gauge potential is introduced in the Hamiltonian through the momentum term by means of momentum translation, $e.g., k_y\rightarrow k_y+A_y(t)$ for a static field along $\hat{y}$. This led us to find evolved wave functions, which are used to compute expectations of different spin operators at an instant of time and for a finite $\varepsilon_\textrm{F}$. Figures~\ref{fig:spin_textureROC}(b) depicts the shifts of FCs under the applied static field along $y$, while the snapshots of spins at an instant of time capture the spin polarisation on the contours. It is evident that the FCs move oppositely away from the direction of the applied field. Furthermore, in comparison to the \textit{spin-momentum} locking in Fig.~\ref{fig:spin_textureROC}(a)(inset), the time-dependent spins ($i.e., \bra{\psi} \boldsymbol{\sigma} \ket{\psi}_k$) in the $k_x$-$k_y$ plane on both the FCs orient at arbitrary directions during the short-time evolution. 
Consequently, the magnitudes of resultant induced polarisation involving both inner and outer FCs, $\langle S_{x(y)}\rangle=\langle S_{x(y)}\rangle_{\rm in}+\langle S_{x(y)}\rangle_{\rm out}$, oscillate initially and then gradually saturates (reaches to a nearly constant value) over time. The converged final value is smaller than the actual value of the outer FC since the spin polarisations of the inner FC are opposite and smaller in magnitude. Figures~\ref{fig:spin_textureROC}(c) and (d) illustrate these features, showing the variation of saturated spin polarisation $\langle S_{y(x)}\rangle$ as a function of time for the field applied along $x (y)$. In this approach, it is evident that the magnitudes of both the spin components, $\langle S_{x}\rangle$ and $\langle S_{y}\rangle$ eventually saturate in the long-time limit. In Fig.~\ref{fig:spin_textureROC} (f), we have computed the time-averaged saturation value of $\langle S_y\rangle$ as a function of the applied field, $\textbf{E}=(E_x, 0, 0)$. Note that the spin polarisation is linear in the weak field limit, corroborating the result obtained from the Kubo formalism in Eq.~(\ref{eq:spin_Rashba}). However, its field-dependent behavior exhibits a maximum at a certain field, undergoing significant changes for stronger fields. For simplicity and without loss of generality, we restrict to linear response regime for the rest of this article. 

The concentric isotropic FCs, characteristic of the linear Rashba effect, are then manipulated by an applied electric field $\textbf{E} =(E\,\cos\theta,E\,\sin\theta,0)$, confining within the plane of a 2DEG. Here, $\theta$ denotes the angle of the static field defined with respect to the 2DEG coordinates, as indicated schematically in Fig.~\ref{fig:spin_textureROC}(a). We note that similar concentric FCs had also been demonstrated at the surface of SrTiO$_3$~\cite{STO_NatMat2011}.
Figures~\ref{fig:spin_textureElectric} (a) and (b) demonstrate the saturated polarisation of $\langle S_x\rangle$ and $\langle S_y\rangle$ as a function of $\theta$ for the inner (1) and outer (2) FCs, respectively. The magnitude of both $\langle S_x\rangle$ and $\langle S_y\rangle$ for both the FCs oscillates with $\theta$.  
However, the total spin polarisation, defined as $\langle S\rangle=\sqrt{(\langle S^{in}_{x}\rangle + \langle S^{out}_{x}\rangle)^2 + \langle S^{in}_{y}\rangle + \langle S^{out}_{y}\rangle)^2}$, remains constant, independent of $\theta$. Here $in$ and $out$ refer to the inner and outer FCs, respectively. The corresponding plot is shown in Fig.~\ref{fig:spin_textureElectric}(c). In Fig.~\ref{fig:spin_textureElectric}(d), we illustrate the direction of spin polarisation with $\textbf{E}$ in the $E_x$-$E_y$ plane. The polarisation turns out to be transverse to the direction of the applied field, corroborating the results of standard linear REE. For a field along any arbitrary $\theta$ value, both the spin components may exist; however, the total induced spin polarisation remains constant in magnitude. Unless explicitly mentioned otherwise, the induced spin polarisations are discussed at their saturation values (long-time limit) in the remaining sections of the article. 

An important characteristic of the RSOC is its controllability through an external gate voltage applied to the 2DEG~\cite{Manchon_2015, Michiardi_2022}. The Rashba parameter is directly influenced by the surface/interfacial potential drop, and adjusting the electron occupation can be achieved by applying a gate voltage~\cite{REE-GateVoltage_PRL1997}, light~\cite{STO_NatMat2011} $etc$.
It is also evident from the Fig.~\ref{fig:spin_textureElectric}(e), the size of these two concentric FCs can be controlled by varying the energy through doping. The energy at the zero-momentum point, where the two Rashba bands intersect, is set to $\varepsilon_0=0$. The point is represented as the band crossing point (BCP). We, therefore, focus on exploring how spin polarisation varies with $\varepsilon_\textrm{F}$ for a given electric field along $\hat{x}$. At $\varepsilon_\textrm{F} = \varepsilon_0$, the induced polarisation originates exclusively from the outermost FC since there is no inner FC. This closely resembles the FC on the surface state of topological insulators above the Dirac point, such as the elemental topological insulator $\alpha$-Sn~\cite{REE-TI_PRL2016}. The resulting spin polarisation $\langle S\rangle$ as a function of $\varepsilon_\textrm{F}$ is shown in Fig.~\ref{fig:spin_textureElectric}(f). Unlike a topological insulator, as $\varepsilon_\textrm{F}$ increases ($> 0$) away from the BCP point, the spin polarisation $\langle S \rangle$ gradually diminishes to zero.
This reduction is attributed to the partial compensation of spin polarisation from carriers with opposite spin chirality in the inner and outer FCs above the BCP point. 
Conversely, a decrease in $\varepsilon_\textrm{F} (< 0)$ for carriers with energy below the BCP leads to an enhancement of $\langle S \rangle$ due to the cooperative contributions of the identical spin helicities from the outer bands.

\section{\label{sec:level3}Spin polarisation in surfaces with linear and non-linear  terms under different symmetries}
The splitting of energy bands induced by SOC and the associated spin-momentum locking, originating from the effective magnetic field within a Rashba Hamiltonian (cf. Eq.~\ref{eq:ham0}), often deviate in systems with bulk-inversion asymmetry. In a typical linear SOC limit with BIA, DSOC emerges, and the interplay between these two SOC contributions is observed in a wide class of noncentrosymmetric materials~\cite{PRL2005, Weber_PRL2007, Feng_NatCommun2019}. This leads to anisotropic FCs where the spin-momentum locking may no longer remain orthogonal~\cite{Nagaosa_PRB2017, PRB_2015}. Moving into the nonlinear $k$-dependent spin splitting, notable examples such as semiconductor quantum wells~\cite{moriya2014, PRL2018}, oxide surfaces, and interfaces~\cite{PRL2012, NatCommun2014} exhibit the presence of cubic Rashba terms, which manifest in various forms within the Hamiltonian.
Remarkably, the effective momentum-dependent magnetic field discussed before takes a distinct form in the presence of cubic SOC terms than its linear counterpart. This enables the $\textbf{k}$-dependent \textit{pseudospin} in those systems to undergo faster rotation around the FCs and the \textit{pseudospin} vector is no longer orthogonal to $\textbf{k}$ everywhere~\cite{PRL2012, PRB2016,usachov2020}. 
The presence of these terms in the Hamiltonian can result in diverse and unconventional field-induced spin polarisations. Surprisingly, despite their potential applications in spintronics and valleytronics, the comprehensive understanding of how the various linear and non-linear-in-$k$ SOC, including DSOC, determine the spin polarisation pattern has been largely overlooked. Therefore, in the next, we focus on the theoretical investigations of the Hamiltonian in Eq.~\ref{eq:ham0} in the presence of linear-in-$k$ dependent DSOC terms allowed by various symmetries such as $C_{3}, C_{4}, C_{3v}$ and $C_{4v}$ symmetries~\cite{vajna2012higher}. 
Thereafter, we investigate Eq.~\ref{eq:ham0} in the presence of higher-order SOC terms, particularly the symmetry-allowed cubic RSOC and DSOC terms. We note that these SOC terms in the Hamiltonian are primarily adopted from Ref.~\cite{vajna2012higher}.

\subsection{Interplay between RSOC and DSOC within generalize linear regime}\label{subRSOCDSCOlin}
We now construct quantum well models in the presence of SIA, BIA, or a combination of both, aiming to investigate current-induced spin polarisation in 2DEG. The model is built upon a symmetry-adapted linear-in-$k$ approximation for the SOC effect. The simultaneous presence of SIA and BIA occurs in lower point-group symmetries, allowing linear terms such as $\beta(k_x\sigma_x+k_y\sigma_y)$, $\beta(k_x\sigma_x-k_y\sigma_y)$, etc.~\cite{vajna2012higher} in the Hamiltonian. The first term is referred to as DSOC($+$), while the second is labeled DSOC($-$) and is commonly known as standard Dresselhaus term. Note that each of the individual linear terms considered (without RSOC) may exhibit distinct ST on the FCs, and subsequently, they show longitudinal spin polarisation feature with respect to the applied field with an opposite sign (see Appendix). This occurs because, even though the eigenvalues of both Hamiltonians are identical, the wavefunctions differ only by phase. Consequently, we observe distinct behavior in the spin polarisation phenomena in the linear regime due to the interplay between the RSOC parameter $\alpha$ and DSOC parameter $\beta$.

\begin{figure*}
\centering
\includegraphics[width=1\linewidth]{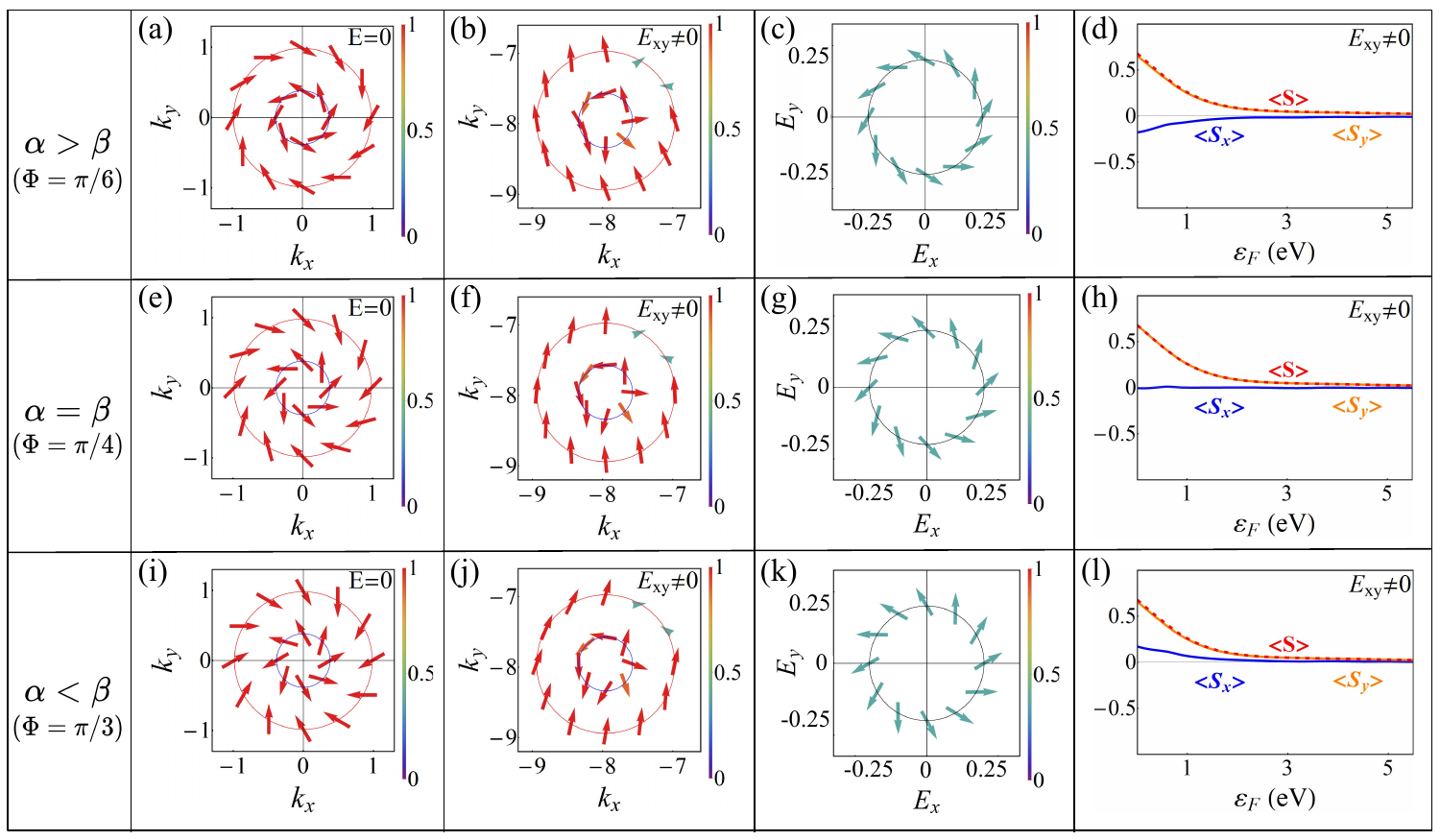}
\caption{Spin orientations of the 2DEG together with RSOC and DSOC(+) for their different relative strengths $\alpha>\beta$, $\alpha=\beta$, and $\alpha<\beta$, respectively (considering $\alpha^2+\beta^2=1$). (a, e, i) Plot of zero field STs. The variation of STs from Rashba-like to radial Weyl-like is evident. (b, f, j) Snapshot of spin orientation for the field along $xy$ at constant energy surface $\varepsilon_\textrm{F}=0.63$ eV and at a particular instant of time ($t=45$~s). (c,g,k)  The resultant time and momentum averaged spin polarisation in the $E_x-E_y$ plane for both FCs at
$\varepsilon_\textrm{F}= 0.63$ eV. (d,h,l)  Variation of spin polarisation per unit area with the Fermi energy.}
\label{fig:DSOCplus+RSOC}
\end{figure*}

\subsubsection{RSOC~+~DSOC(+):} We first focus on the DSOC($+$) without and with the RSOC term within the linear-in-$k$ SOC limit. The full Hamiltonian reads of
\begin{align}
H=\frac{\hbar^2 k^2}{2 m}+\alpha(k_x \sigma_y -k_y \sigma_x) +\beta(k_x\sigma_x+k_y \sigma_y).
\label{eq:rashba_dress+}
\end{align}
The Hamiltonian in Eq.~\ref{eq:rashba_dress+} characterizes a 2DEG without mirror symmetry under point group $C_3$ or $C_4$~\cite{vajna2012higher}.
For simplicity, we relate the coefficients $\alpha$ and $\beta$ by defining an angle $\Phi = \tan^{-1}(\beta/\alpha)$, while maintaining $\alpha^2 + \beta^2 = 1$. In standard Rashba-Dresselhaus systems, the STs varies depending on the value of $\Phi$; $\Phi$ = 0 and $\pi/2$ represent pure Rashba and Dresselhaus systems, respectively. For the later case, diagonalizing only the DSOC($+$) term yields radial STs, where the spin is always aligned parallel to the wave vector $\textbf{k} = (k_x, k_y, 0)$, historically referred to as the \textit{Weyl}-type STs~\cite{Tao, Weyl1929, krieger2024weyl, Veneri}. This type of ST has recently been realized in chiral bulk crystals, where both mirror and inversion symmetries are absent~\cite{ChiralCrys_PRL2020-1, ChiralCrys_PRL2020-2}. 
When subjected to an applied electric field, this term generates spin polarisation parallel to the direction of the electric field, contrasting sharply with the pure Rashba systems (associated with $C_{3v}$ or $C_{4v}$ point group), where the induced spin polarisation is perpendicular to the field. The ability to tune the $\alpha$ and $\beta$ parameters is therefore expected to manifest in a controlled ST within the FCs, thereby influencing the spin polarisation induced by an electric field.

As before, using the Kubo formalism, the induced spin polarisation components for $\textbf{E}=(0, E_y,0)$ are found to be (see Appendix~\ref{app:kubo1} for detailed derivation)
\begin{align}
    \langle S_x\rangle&\approx -\frac{e E_y \alpha \varepsilon_F}{32\pi\eta(\alpha^2+\beta^2)};\nonumber\\
    \langle S_y\rangle&\approx \frac{e E_y \beta \varepsilon_F}{32\pi\eta (\alpha^2+\beta^2)};~~\nonumber\\
    \langle S_z\rangle&=0.~~~~~~~~~~~~~~~~~~~~~~~~
    \label{eq6}
\end{align}
Clearly, upon looking at each spin polarisation component, it is evident that the transverse component $\langle S_x\rangle$ varies with $\alpha$, while the longitudinal component $\langle S_y\rangle$ varies with $\beta$ but with an opposite sign.
In contrast, for the field $\textbf{E}=(E_x,0,0)$, we find that both $\langle S_x\rangle$ and $\langle S_y\rangle$ have the same sign, as expressed by:
\begin{align}
\langle S_x\rangle&\approx \frac{e E_x \beta \varepsilon_F}{32\pi\eta (\alpha^2+\beta^2)};\nonumber \\ 
\langle S_y\rangle&\approx\frac{e E_x \alpha \varepsilon_F}{32\pi\eta (\alpha^2+\beta^2)};\nonumber\\
\langle S_z\rangle&=0.~~~~~~~~~~~~~~~~~~~~~~
\label{eq7}
\end{align}
Such contrasting behavior of $\langle S_x\rangle$ and $\langle S_y\rangle$ for the fields along $\hat{x}$ and $\hat{y}$ directions can be understood by the effective momentum-dependent magnetic field $
\textbf{B}_\textrm{eff} (\textbf{k}) = (\beta k_x-\alpha k_y, \alpha k_x+\beta k_y)$ acting on spin $\boldsymbol{\sigma}$. The sign of $k_y$ in the $x$ and $y$ components of $\textbf{B}_{\rm eff}$ differs. Accordingly, the external field along $\hat{y}$ couples to both $k_y$ via Peierls substitution, resulting in spin polarisation along $\hat{x}$ and $\hat{y}$ directions with a different sign. However, the field along $\hat{x}$ produces the same sign of $\langle S_x\rangle$ and $\langle S_y\rangle$ since the sign of $k_x$ is the same in both the components of $\textbf{B}_{\rm eff}$. Note that this is one of the {\it crucial} results which has been overlooked in the existing literature~\cite{schliemann2003anisotropic,schliemann2003nonballistic}. The orientation of spins around the FCs is influenced by the sign change in $\langle S_x\rangle$ and $\langle S_y\rangle$, as will be detailed shortly.
\begin{figure*}
\centering
\includegraphics[width=1\linewidth]{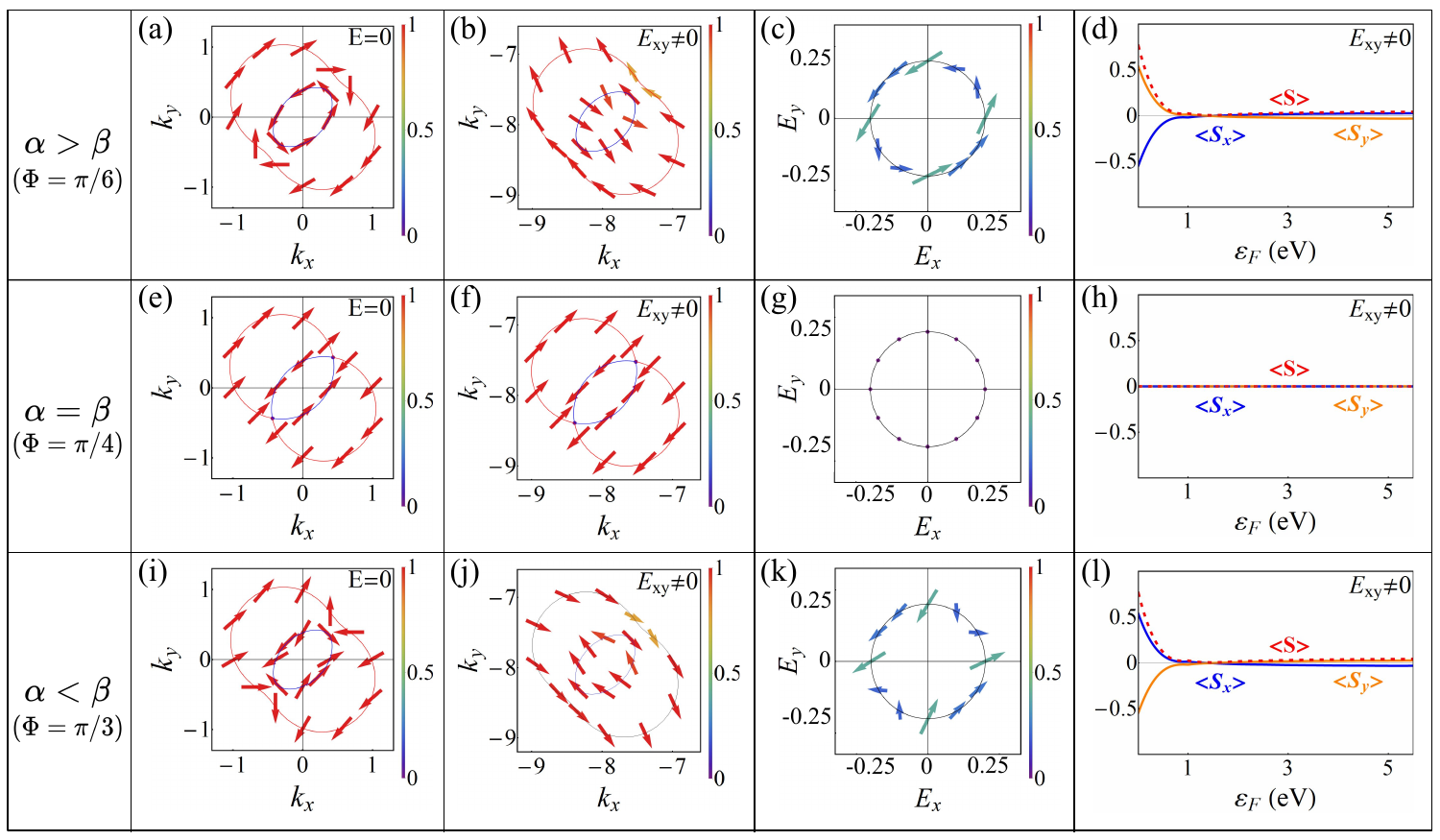}
\caption{Same as Fig.~\ref{fig:DSOCplus+RSOC} for the 2DEG together with RSOC and DSOC($-$). Note that the spin polarisation here differs significantly from the DSOC(+) case, as elaborated in the main text. Additionally, the $\alpha=\beta$ case is even more interesting as it shows persistent ST. Consequently, the net-induced spin polarisation becomes zero.}
\label{fig:DSOCmnus+RSOC}
\end{figure*}

To corroborate the result of Kubo formalism, we solve the time-dependent Schr\"odinger equation using the Hamiltonian in Eq.~(\ref{eq:rashba_dress+}) and subsequently, we compute the spin polarisation after the time evolution. The STs for different values of $\Phi$ (= $\pi/6$, $\pi/4$, and $\pi/3$) is shown in Fig.~\ref{fig:DSOCplus+RSOC}(a),(e) and (i) at zero electric field. The FCs remain isotropic and circular regardless of the values of $\alpha$ and $\beta$.
However, it is evident that the ST around both FCs varies, ranging from the conventional tangential Rashba STs ($\beta=0$ case) to radial Weyl-type STs ($\alpha=0$ case). This interplay, which governs controlled STs, could provide a productive foundation for designing compounds with specific target STs, consequently allowing precise control over the induced spin polarisation vector.
For example, both the spin polarisation $\langle S_x \rangle$ and $\langle S_y \rangle$ are finite for finite $\alpha$ and $\beta$, regardless of the direction of the applied field. 
Accordingly, in response to an electric field perturbation ($e.g.$, $\mathbf{E}_{xy} \ne 0$), the orientation of spins in the ST starts to adjust, deviating from their initial direction as observed in the $k_x$-$k_y$ plane, see FCs in Figs.~\ref{fig:DSOCplus+RSOC}(b), (f) and (j). After a sufficiently long time, the momentum-dependent pseudospins lead to an induced moment, and the orientation of its vector varies with the change in the direction of the applied electric field. This is illustrated in Figs.~\ref{fig:DSOCplus+RSOC}(c), (g), and (k), which show the induced in-plane spin moment vector $\mathbf{S}=\langle S_x \rangle \hat x+\langle S_y\rangle\hat y$ as a function of the direction of an in-plane $\mathbf{E}$. 
We observe orthogonal Rashba-like and radial Weyl-like ST in the induced polarisation concerning the direction of $\mathbf{E}$, corresponding to $\Phi$ values of 0 and $\pi/2$, respectively (not shown here). The orientation of the spin moment, given by ($\langle S_x \rangle$, $\langle S_y \rangle$, 0), varies with the direction of $\mathbf{E}$ in the $xy$-plane, aligns with the analytical predictions derived from the Kubo formula. Regardless of the direction of the field, we consistently observe the rotation of induced moment while its magnitude remains constant, as indicated by different $\Phi$ values. This directly demonstrates that the electrically induced polarisation depends on the underlying spin texture itself. In Fig~\ref{fig:DSOCplus+RSOC}(d),(h),(l), we illustrate how the induced polarisation changes as the Fermi level varies. As discussed earlier, the magnitude of the spin polarisation gradually diminishes.

\subsubsection{RSOC~+~DSOC($-$):}
In the linear-in-$k$ SOC Hamiltonian (see Eq.~\ref{eq:rashba_dress+}), it is worth noting that the Dresselhaus term does not have a unique expression. The derivation of $k$-dependent SOC terms typically arises from zone-center solutions of the $\mathbf{k.p}$ theory~\cite{HigherOrder_PRB2006}.
The Hamiltonian incorporating standard Dresselhaus term is expressed as:
\begin{align}
H=\frac{\hbar^2 k^2}{2 m}+\alpha(k_x \sigma_y -k_y \sigma_x) +\beta(k_x\sigma_x-k_y \sigma_y).
\label{eq:rashba_dress-}
\end{align}
Such a Hamiltonian can describe the interfacial SOC effect in III-V semiconductor quantum wells~\cite{KohdaPRB2012}. 
Diagonalizing standard Dresselhaus term, denoted as DSOC($-$), we obtain \textit{tangential-radial} spin orientation on FCs, commonly known as the Dresselhaus STs. Surprisingly, the findings outlined in the preceding section exhibit substantial distinctions from the outcomes obtained by analyzing the electric field response for the Hamiltonian in Eq.~\ref{eq:rashba_dress-}. Results are summarized in Fig.~\ref{fig:DSOCmnus+RSOC}. 
Such contrasting results, even within the linear-in-$k$ SOC limit, arise from the specific design of the 2DEG interface systems. 
In this scenario, the FC no longer remain circular; instead, there is a lateral shift along the diagonal of $k_x$-$k_y$ plane. We note here that the anisotropic spectrum in the ST complicates the derivation of simple expressions for $\langle S_x\rangle$ and $\langle S_y\rangle$. However, the sign of $\langle S_x\rangle$ and $\langle S_y\rangle$ can be determined using the effective magnetic field $\mathbf{B}_{\rm eff}=(\beta k_x-\alpha k_y,\alpha k_x-\beta k_y)$. Since both the components of $\mathbf{B}_{\rm eff}$ are having a similar sign with $k_x$ and $k_y$, the sign of $\langle S_x\rangle$ and $\langle S_y\rangle$ are expected to be same.

The induced spin polarisation shown in Fig.~\ref{fig:DSOCmnus+RSOC} for $\Phi = \pi/6$ and $\pi/3$, respectively (Fig.~\ref{fig:DSOCmnus+RSOC} c and k), varies in magnitude depending on the direction of the electric field. Evidently, the ST on the FCs differs substantially from those discussed in the previous DSOC(+) case. Additionally, the asymmetric nature of the FCs leads to a more rapid decay of spin polarisation to zero with increasing $\varepsilon_\textrm{F}$ shift compared to the DSOC(+) case (see Fig.~\ref{fig:DSOCmnus+RSOC}(d) and (l)).   
In the present case, a special situation arises when the Rashba and Dresselhaus parameters in the Hamiltonian are of equal magnitude at $\Phi=\frac{\pi}{4}$ ($\alpha = \beta$). The ST remains invariant with momentum as the effective magnetic field $\mathbf{B}_{\rm eff}= (\alpha(k_x-k_y),\alpha(k_x-k_y))$ becomes a constant vector, leading to a persistent spin texture (PST) with unidirectional spin alignment~\cite{Tao_NatCommun2018}. This is evident from the time evolution of pseudospins on FCs, compare STs in  Figs.~\ref{fig:DSOCmnus+RSOC}(e) and (f). Consequently, the system does not respond to the applied field, resulting in zero spin polarisation as shown in \ref{fig:DSOCmnus+RSOC}(h).

\begin{figure}
\centering
\includegraphics[width=0.69\linewidth]{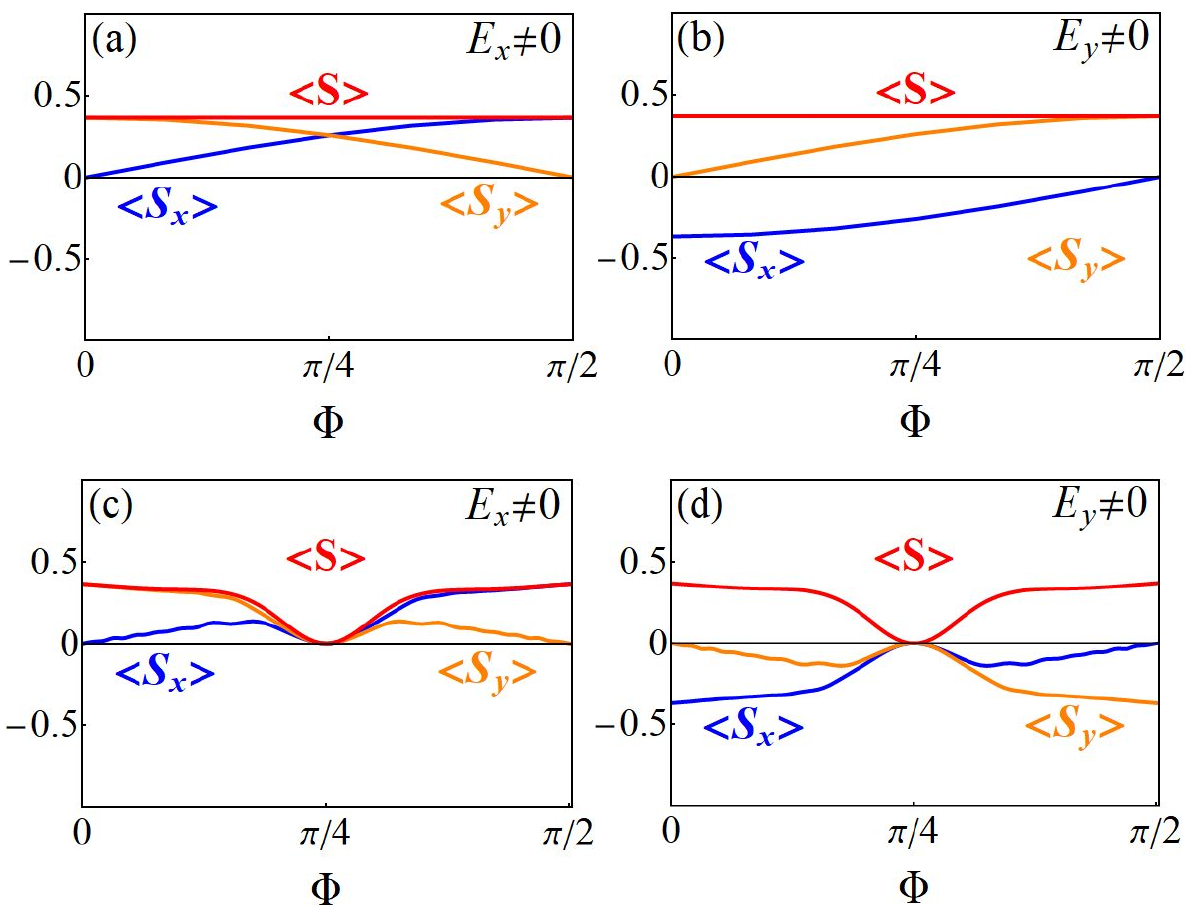}
\caption{Induced spin polarisation (at $\varepsilon_\textrm{F}=0.63$ eV) as a function of $\Phi$ = tan$^{-1}(\beta/\alpha)$. (a,b) and (c,d) illustrate the interplay between the relative strengths of the DSOC(+) and DSOC($-$) terms with respect to the RSOC term, respectively.}
\label{fig:alpha_betaRatio}
\end{figure}

An earlier study demonstrated anisotropic transport in semiconductor quantum wells that confine electron gas under the influence of both the RSOC and DSOC($-$) term. This anisotropy is determined by the interplay between the $\alpha$ and $\beta$ parameters~\cite{schliemann2003anisotropic}. We highlight that the spin polarisation obtained within the REE framework, considering the presence of RSOC and standard DSOC($-$), exhibits anisotropy in induced polarisation. The contrasting behavior of the DSOC($+$) and DSOC($-$) terms in response to an electric field is illustrated in Fig.~\ref{fig:alpha_betaRatio}, as a function of $\Phi$. For a given applied field along $\theta = 0$ and $\pi/2$, the magnitude of induced polarisation ($|\mathbf{S}|$) remains almost constant with $\Phi$ in the case of DSOC($+$), while it gradually decreases to zero for DSOC($-$) at $\Phi=\pi/4$, i. e., at the regime of PST. Although the PST has been acknowledged for its vital role in establishing an infinite spin lifetime~\cite{Bernevig}, it is worth noting that the induced spin polarisation vanishes in the presence of the electric field. 

\begin{figure}[t!]
\centering
\includegraphics[width=0.66\linewidth]{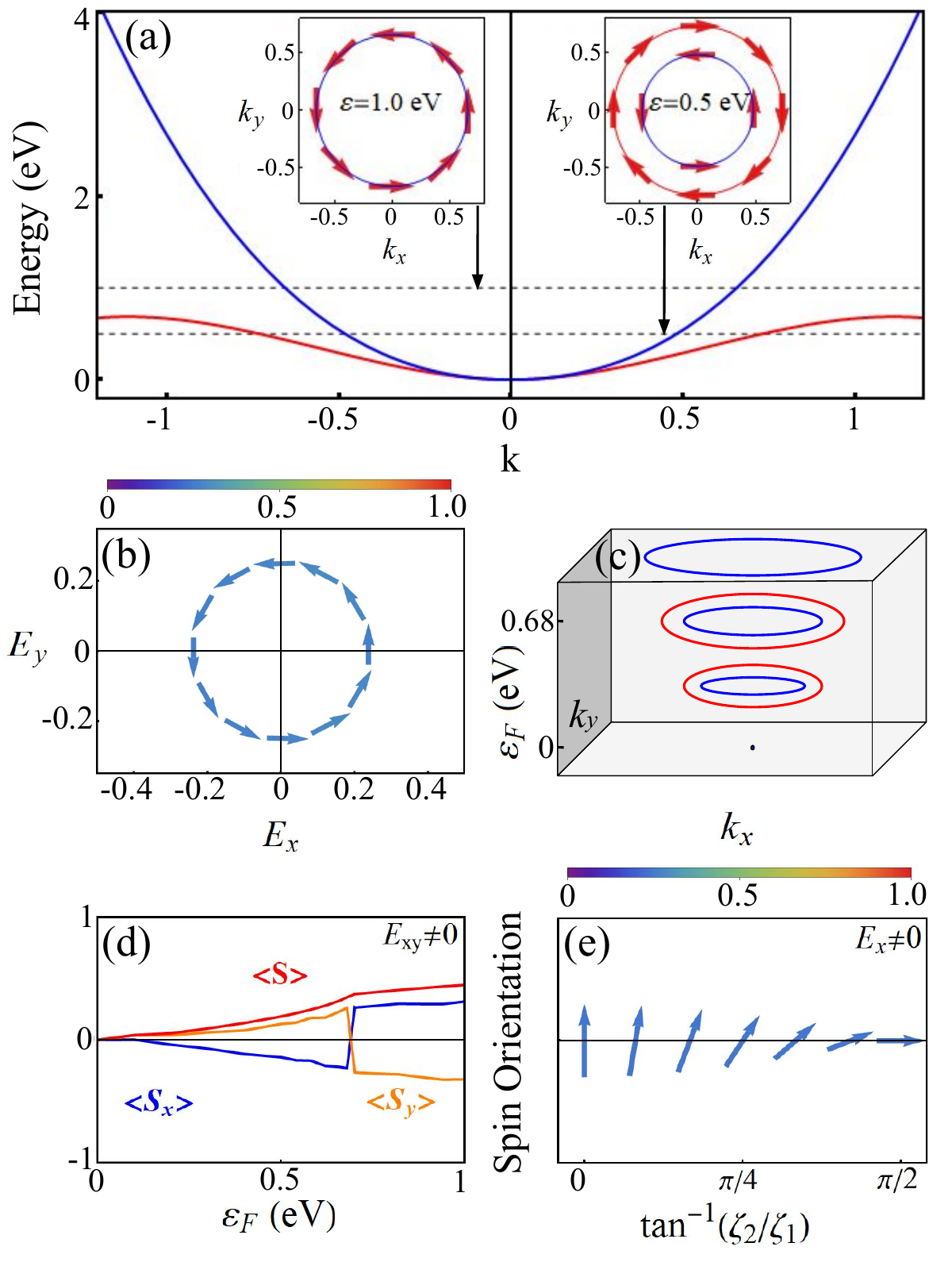}
\caption{(a) Low-energy band spectrum of Eq.~(\ref{eq:only_cubic}). The insets show spin orientations along the FCs at two different energies $\epsilon_\textrm{F}=1.0$ eV and $0.5$ eV, respectively. (b) Spin polarisation per unit area in the $E_x$-$E_y$ plane. Clearly, it is transverse to the applied field similar to standard REE. (c) and (d) Variation in the shape of FCs and spin polarisation with $\varepsilon_\textrm{F}$, respectively. Evidently, the polarisation enhances with the Fermi energy as opposed to standard RSOC term. (e) The orientation of induced spin polarisation as a function of relative strength between cubic RSOC and cubic DSOC under $C_3$ symmetry.}
\label{Cubic RSOC C3}
\end{figure}

\subsection{\label{sec:level4} Non-linear regime: the symmetry allowed cubic order effect}\label{subRSOCDSOCnonlin}
The orientation of spins with momentum in a more complex ST, governed by higher-order SOC terms, enables the emergence of exotic current responses in non-centrosymmetric materials. For example, cubic-in-$k^3$ Rashba terms can substantially change the spin polarisation in heavy hole quantum well than the linear-in-$k$ Rashba SOC~\cite{moriya2014,CubicRashbaNakamura_PRL2012}. Although the cubic term, in general, accompanies the linear-in-$k$ term, some experiments reported {\it purely} cubic-in-$k$ Rashba SOC in surfaces/interfaces of materials such as the surface of SrTiO$_3$(001) \cite{CubicRashbaNakamura_PRL2012}, Ge/SiGe quantum well~~\cite{moriya2014}, asymmetric oxide heterostructures like LaAlO$_3$/SrTiO$_3$/LaAlO$_3$~\cite{lin2019}, and surfaces of antiferromagnet TbRh$_2$Si$_2$~\cite{usachov2020}. 
Bulk materials, particularly with $\Bar{6}$ and $\Bar{6}m2$ point groups~\cite{zhao2020PRL}, are also found to exhibit purely cubic SOC term.

\subsubsection{Cubic terms associated with $C_{3v}$ and $C_3$ symmetries}\label{subsubc3v}
The two-band Hamiltonian with cubic Rashba terms associated with $C_{3v}$ point group symmetry~\cite{vajna2012higher} is given by
\begin{align}
H= \frac{\hbar^2k^2}{2 m} +\zeta_1 ((k_x^3+k_x k_y^2)\sigma_y -(k_x^2 k_y +k_y^3)\sigma_x),
\label{eq:only_cubic}
\end{align} 
where $\zeta_1$ is the strength of the cubic SOC. This higher-order RSOC usually manifests in [111]-oriented diamond and zinc-blende quantum wells, as well as in their surface states~\cite{HigherOrder_PRB2006}. The eigenvalues of Eq.~(\ref{eq:only_cubic}) can be easily obtained as $E_\pm$=$\hbar^2 k^2/2m \pm \zeta_1 k^3$ and they are plotted in Fig.~\ref{Cubic RSOC C3}(a) over the range of $\pm|\mathbf{k}|$ in the ($k_x, k_y$) plane. To explore the Edelstein phenomena in this 2DEG setup, we ensure that the kinetic energy consistently dominates over the energy associated with the SOC term. Considering the concepts of the effective magnetic field discussed before, the presence of field $\textbf{E} = (E_x, 0, 0)$ is expected to induce polarisation in both the transverse and longitudinal directions of the applied field as both the Pauli matrices involves $k_x$ and $k_y$ in the Hamiltonian. 
Remarkably, Fig.~\ref{Cubic RSOC C3}(b) shows that the pure cubic RSOC results in transverse Polarisation, akin to the behavior exhibited by linear RSOC coupling in Fig.~\ref{fig:spin_textureElectric}(d). This is because the effective magnetic field can be expressed as $\mathbf{B}_\textrm{eff}(\mathbf{k}) =\zeta_1 k^3(-\sin\delta,~\cos\delta)$ which resembles the effective magnetic field for the linear-in-k case, $\mathbf{B}_\textrm{eff}(\mathbf{k}) =\alpha k (-\sin\delta,~\cos\delta)$, where $\delta$ is the polar angle. The associated spin splitting turns out to be proportional to $k^3$ and isotropic (independent of $\delta$), yielding concentric rings in the constant energy cut. The STs shown in the inset of Fig.~\ref{Cubic RSOC C3}(a) for energy $\varepsilon \ne 0$ also closely resembles that of the linear-in-$k$ SOC Hamiltonian. 
 
In Figure~\ref{Cubic RSOC C3}(d), we depict the dependence of the net spin polarisation $\langle S \rangle$ alongside its in-plane components, $\langle S_x \rangle$ and $\langle S_y \rangle$, with respect to the tunable Fermi energy, $\varepsilon_\textrm{F}$. Contrary to the behavior in the case of linear-in-$k$ Rashba
SOC (see Fig.~\ref{fig:spin_textureElectric}(f)), the magnitude of spin polarisation $\langle S \rangle$ increases with the increase in $\varepsilon_\textrm{F}$.
This is attributed to the chiral spin texture in a single FC for  $\varepsilon_\textrm{F}>0.68$ eV (Fig.~\ref{Cubic RSOC C3}(c)). This strikingly resembles the phenomena arising from a single chiral surface state loop in a topological insulator~\cite{TI_PRL2010}. Thus, in a more general context, the higher-order term acts as a tool to counteract the cancellation of current-induced spin density observed in standard linear REE. 

When mirror symmetry is broken, such as in the  $C_3$ point group, a cubic-in-$k$ Dresselhaus term, $\zeta_2 ((k_x^3+k_x k_y^2)\sigma_x + (k_x^2 k_y +k_y^3)\sigma_y)$, combines with the Hamiltonian in Eq.~\ref{eq:only_cubic}. Then the current induced spin phenomena is found to depend on the relative values of $\zeta_1$ and $\zeta_2$. Consequently, the orientation of the induced spin polarisation rotates for a specific applied field direction, similar to the case of the linear-in-$k$ Rashba-Dresselhaus Hamiltonian. The orientation of the induced spin polarisation vector is illustrated in Fig.~\ref{Cubic RSOC C3}(e) for a constant electric field direction $\textbf{E}=(E_x,0,0)$. It varies depending on their relative strengths defined by  tan$^{-1}(\zeta_2/\zeta_1)$. The induced polarisation vector continues to rotate from Rashba-type ($\perp \textbf{E}$) to Dresselhaus-type ($||~\textbf{E}$), reflecting the interplay between RSOC and DSOC.

\begin{figure}
\centering
\includegraphics[width=0.66\linewidth]{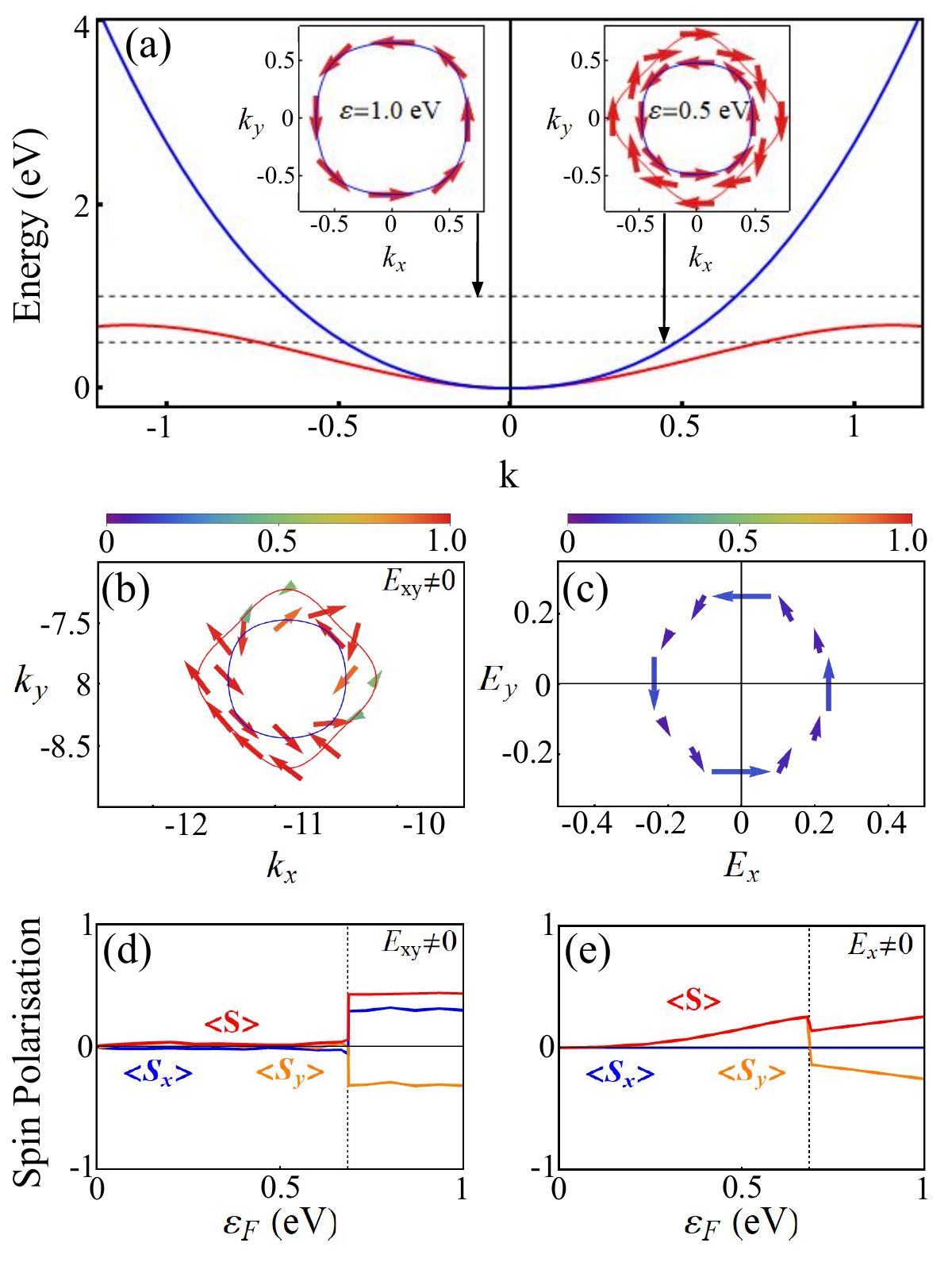}
\caption{(a) The energy dispersion and spin orientations along the FCs of constant energy $\varepsilon_\textrm{F}=0.5$ and $1.0$ eV, respectively (in the inset) for pure-$k^3$ RSOC term under $C_{4v}$ symmetry. (b) Snapshot of orientations of spins when electric field applied along $xy$-direction. (c) Resultant time and momentum averaged spin Polarisation as a function of electric field in the $E_x$-$E_y$ plane for both the FCs at $\varepsilon_\textrm{F} = 0.5$ eV. (d, e) Spin polarisation as a function of $\varepsilon_\textrm{F}$ for the electric field along $xy$- and $x$-direction, respectively. The vertical line refers to the ($\varepsilon_\textrm{F}=0.685$ eV) at which we see two FCs to a single FC transition.}
\label{Cubic RSOC C4 2nd term}
\end{figure}

\subsubsection{Cubic terms associated with $C_{4v}$ and $C_4$ symmetries}\label{subsubc4v}

\begin{figure}
\centering
\includegraphics[width=0.66\linewidth]{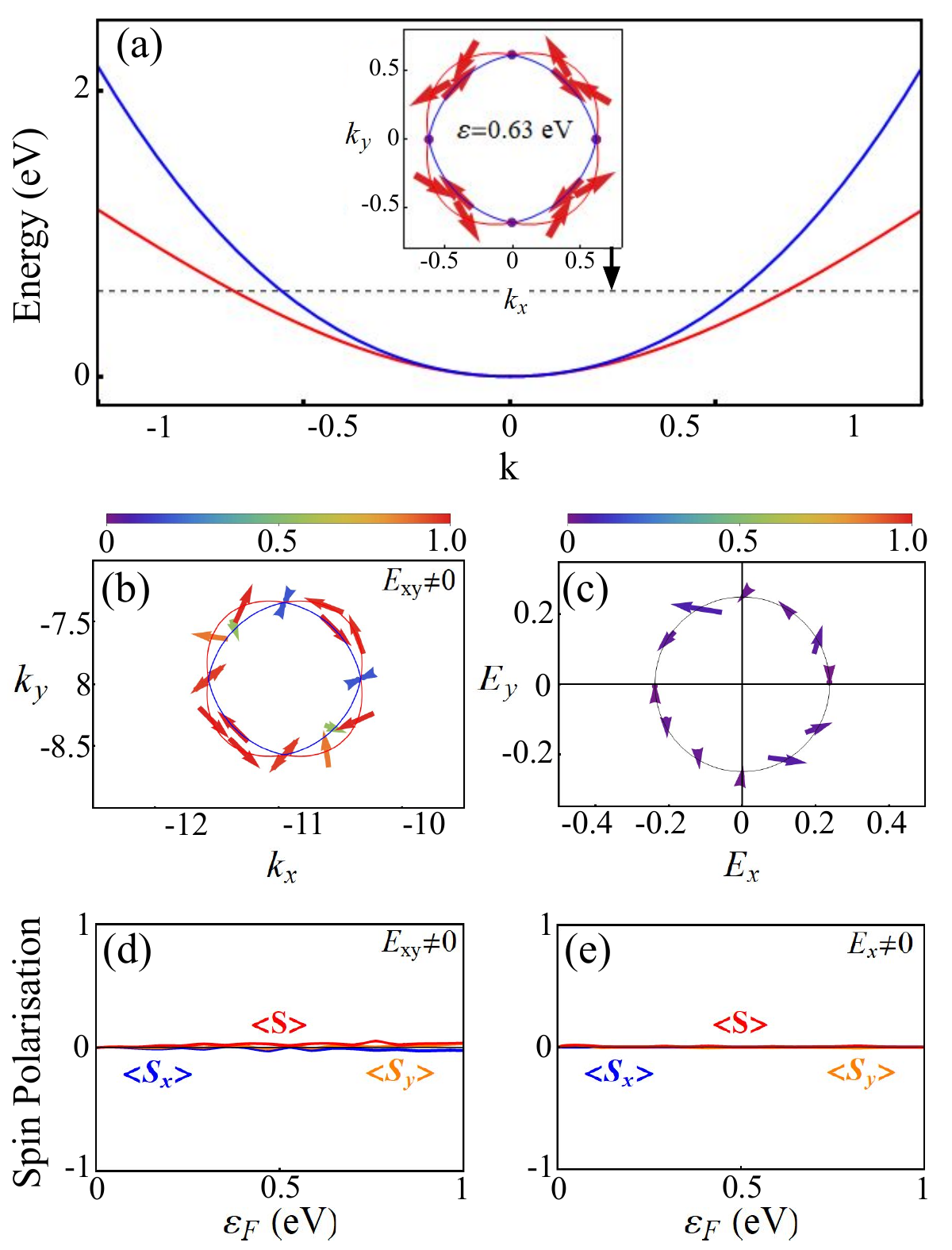}
\caption{(a) The energy dispersion and spin orientations along the FCs of constant energy $\varepsilon_\textrm{F}=0.6$~eV (inset) for the allowed mixed-$k^3$ RSOC term, $\gamma_1(k_x^2k_y\sigma_x-k_xk_y^2\sigma_y)$, under $C_{4v}$ symmetry. b) The snapshot of the spin polarisation in the $k_x$-$k_y$ plane when the electric field is applied diagonally, the $xy$-direction. c) The electric field direction dependency of the spin polarisation in $E_x$-$E_y$ plane (at constant energy $\varepsilon_\textrm{F}=0.63$ eV). (d, e) Spin polarisation with $\varepsilon_\textrm{F}$ for the field applied diagonally along $xy$- and $x$-direction, respectively.}
\label{Cubic RSOC C4 3rd term}
\end{figure}

We next move to another class of point groups, $i.e.$, $C_{4v}$ and $C_{4}$, considering a quantum well geometry of 2DEG. For $C_{4v}$, only the Rashba splitting is expected, with the allowed cubic-in-$k$ RSOC terms being $\lambda_1(k_x^3\sigma_y-k_y^3\sigma_x)$ and $\gamma_1(k_x^2k_y\sigma_x-k_xk_y^2\sigma_y)$, representing pure-$k^3$ and mixed-$k^3$ terms, respectively~\cite{vajna2012higher}. This higher-order RSOC often characterizes the Hamiltonian of surface states in diamond and zinc-blende quantum wells oriented along the [001] direction~\cite{HigherOrder_PRB2006}. The eigenvalues corresponding to the pure-$k^3$ and mixed-$k^3$ SOC are given by $E_\pm$=$\frac{\hbar^2 k^2}{2m} \pm \lambda_1 \sqrt{k_x^6 + k_y^6}$ and $E_\pm$=$\frac{\hbar^2 k^2}{2m} \pm \gamma_1 k k_x k_y$, respectively. Fig.~\ref{Cubic RSOC C4 2nd term}(a) and Fig.~\ref{Cubic RSOC C4 3rd term}(a) show the respective energy dispersion and $k$-dependent pseudospins on FCs (in the insets) at zero electric fields. The fourfold warping of FCs at low energy is evident in both cases. As before, the non-linear SOC terms involving $\lambda_1$ and $\gamma_1$ result in in-plane effective magnetic fields $\mathbf{B}_\textrm{eff}$ given by $\lambda_1(k^3/4)(-3\sin\delta + \sin 3\delta,~ 3\cos\delta + \cos 3\delta)$ and $\gamma_1(k^3/4)(\sin\delta + \sin 3\delta, -\cos\delta + \cos 3\delta)$, respectively.
Concomitantly, the spin splittings exhibit a dependence on $\delta$, resulting in anisotropic spin orientation along the FCs as evident from the insets of Fig.~\ref{Cubic RSOC C4 2nd term}(a) and Fig.~\ref{Cubic RSOC C4 3rd term}(a). This is in contrast to the $C_{3v}$ case.
Accordingly, the response to the electric field in a 2DEG with $C_{4v}$ symmetry shows contrasting effects. Fig.~\ref{Cubic RSOC C4 2nd term}(c) depicts the induced spin polarisation which exhibits anisotropic behavior, with the maximum value occurring along the longitudinal direction of the applied field. 
In Fig.~\ref{Cubic RSOC C4 2nd term}(d), for the in-plane electric field along xy direction, we observe that the polarisation appears to be zero below $\varepsilon \approx 0.685$~eV due to cancellation from contributions of the two constant energy cuts. Above that energy value, a single FC exists, and it governs a finite net polarisation as shown on the right side of the vertical dashed line in Fig.~\ref{Cubic RSOC C4 2nd term}(d). 
For an electric field along $x$, the polarisation gradually increases until $\varepsilon \approx 0.685$ eV, and a sudden jump in its magnitude at this point can be attributed to the single FC. Following the increase in $\varepsilon_\textrm{F}$, the polarisation magnitude grows gradually. The difference in direction-dependent induced spin polarisation behavior is the consequence of the deformed nature of the FCs; accordingly, the spins oriented perpendicular to the applied field are more responsive than the spins aligned in any other direction.
Interestingly, the mixed-$k^3$ term does not exhibit any generation of spin polarisation even with varying $\varepsilon_\textrm{F}$ (see Fig.~\ref{Cubic RSOC C4 3rd term}(d) and (e)). This absence of induced polarisation retains regardless of the direction of the applied electric field ({Fig.~\ref{Cubic RSOC C4 3rd term}(c)}).

The DSOC terms for $C_4$ point group,  $\lambda_2(k_x^3\sigma_x+k_y^3\sigma_y)$ and $\gamma_2(k_x k_y^2\sigma_x+k_x^2k_y\sigma_y)$~\cite{vajna2012higher}, maintain the warping effect, while the ST changes based on the relative strengths of the $\lambda_2$ and $\gamma_2$ parameters. The induced spin polarisation in these cases also rotates depending on the direction of the applied field, which is influenced by the ratio between the RSOC and DSOC parameters.
Particularly, these terms correspond to DSOC($+$), and similarly, one can identify associated DSOC($-$) terms, analogous to the linear-in-$k$ scenario. It is worth noting that in the pure cubic case, equal strengths of RSOC and DSOC($-$) indeed result in PST generation. This is attributed to the wavefunctions becoming independent of momentum, leading to the pseudospin orientations in STs no longer being dependent on the momentum vector. Therefore, this PST phenomenon is once again observed to be unaffected by the applied field in the time-dependent dynamics, resulting in a net zero-induced spin magnetization. Moreover, similar to the Hamiltonian in Eq.~\ref{eq:rashba_dress-}, the precise tuning of RSOC and DSOC($-$) parameters dictates the overall magnitude of the spin polarisation.
 
\section{\label{sec:level5} Summary and outlook \protect}
We have investigated current-induced spin polarisation, namely the REE effect in 2DEG setup of quantum well geometry, considering all possible linear and nonlinear $k$-dependent Rashba and Dresselhaus SOC terms under various point group symmetries: $C_3$, $C_{3v}$, $C_4$, $C_{4v}$. Possible practical realization can be obtained in non-magnetic noncentrosymmetric bulk and surface/interface materials. Our findings can be summarised as follows: (i) While linear RSOC under $C_{3v}$ or $C_{4v}$ leads to standard REE, the presence of additional linear DSOC due to BIA under $C_{3}$ or $C_{4}$ symmetry leads to distinct features in spin polarisation. In particular, we identify that two symmetry-allowed linear DSOC terms, DSOC(+) and DSOC($-$), have adverse effects on the ST of FCs due to the phase difference in their wavefunctions. These features of STs are manifested in the spin polarisation in the presence of an external field when they are considered individually with the linear RSOC. While RSOC and DSOC(+) lead to finite spin polarisation irrespective of the relative strengths between them, the spin polarisation for RSOC and DSOC($-$) together vanishes when they are equal in strength, leading to persistent spin texture. Additionally, for  DSOC(+) together with RSOC, the spin polarisation remains constant, irrespective of the field direction. In contrast, for the case of DSOC($-$), the magnitude of spin polarisation varies with the field direction.  
In all these cases, we find the spin polarisation to reduce as we increase $\varepsilon_\textrm{F}$. (ii) For nonlinear case, both $C_{3v}$ and $C_{4v}$ allows purely cubic RSOC. Accordingly, we obtain transverse spin polarisation similar to standard REE. However, the magnitude of spin polarisation turns out to vary with the direction of the applied field for $C_{4v}$ in contrast to the $C_{3v}$ case. This fact can be used as a diagnostic measure of the underlying symmetry of metallic surfaces of 2DEG. Additionally, we find that the $C_{4v}$ allows mixed-cubic RSOC, but the spin polarisation, in this case, turns out to be negligibly small due to the deformed FCs. (iii) We further find notable distinctions between linear and cubic RSOC in their spin polarisation. While the spin polarisation reduces as we increase $\varepsilon_\textrm{F}$ for linear RSOC, the cubic RSOC does the opposite due to the presence of a single FC at higher energies. We note that many of these interesting current-induced spin polarisation phenomena between linear and nonlinear RSOC and DSOC have been previously overlooked. Our detailed analyses highlight the intricate and tunable nature of spin polarisation in systems with combined RSOC and DSOC, offering new insights into spintronic applications.   

The idea developed here can be useful for current-induced orbital polarisation, namely the orbital Edelstein effect. Generally, orbital polarisation is subdominant than the spin polarisation~\cite{annika2023}.
However, different Rashba and Dresselhaus parameters can enhance orbital polarisation. In particular, for mixed-$k^3$ Rashba, the system may exhibit only orbital polarisation as the spin polarisation is negligibly small. Further, the interplay between Rashba and Dresselhaus is expected to behave differently depending on the orbital coupling.

\section{Acknowledgement}{We acknowledge financial support from the Department of Atomic Energy (DAE), Govt. of India, through the project Basic Research in Physical and Multidisciplinary Sciences via RIN4001. AD acknowledges the Virgo cluster, where most of the numerical calculations were performed, and thanks to Dr. Arghya Mukherjee for useful discussion and help in the numerical calculations. KS acknowledges funding from the Science and Engineering Research Board (SERB) under SERB-MATRICS Grant No. MTR/2023/000743. }

\appendix*
\section{Calculation of Spin polarisation} \label{app:kubo1}
This appendix provides a detailed calculation of the different components of spin polarisation (Eq. (4), (6), (7) of the main text), resulting from applying an electric field in various directions using the Kubo linear response theory.

We begin with the Hamiltonian for a system that includes a linear k-dependent RSOC and a DSOC(+) term (for simplicity, we neglect the kinetic energy term, as it does not change the eigenfunctions):
\begin{align}
    H= \alpha (k_x \sigma_y -k_y \sigma_x)+\beta (k_x \sigma_x +k_y \sigma_y),
    \label{A1}
\end{align}
where $(k_x,k_y)$ represents the $2D$ crystal momentum, $(\sigma_x,\sigma_y)$ denotes the Pauli matrices, and $\alpha$,$\beta$ corresponds to the Rashba and Dresselhaus coupling parameters, respectively. The energy dispersion becomes,
\begin{align}
    \varepsilon_{\pm}=\pm |k| \sqrt{\alpha^2+\beta^2} .
\end{align}
\vskip 1cm
\subsection{Electric field applied along y direction}
For the electric field along $y$, i. e., $\mathbf{E}=(0,E_y,0)$, the $x$ component of the spin polarisation (from Eq.~\ref{eq3} of the main text) can be expressed as,
\begin{align}
    \langle S_x\rangle = \frac{e E_y}{2\pi}\int \frac{d^2k}{(2\pi)^2} Tr(\hat{v_y} G_A(\varepsilon_F) \hat{S_x} G_R(\varepsilon_F)),
    \label{A3}
\end{align}
where the $y$ component of the velocity operator $\hat{v_y}$ is given by
\begin{align}
    v_y= \begin{pmatrix}
   0 & -\alpha-i\beta\\
-\alpha+i\beta& 0
\end{pmatrix}.
\end{align}
Here $G_R$ and $G_A$ are the retarded and advanced Green's functions, respectively. They are expressed as

\begin{align}
    G_{R/A}= \frac{1}{(\alpha^2+\beta^2)k^2-(\varepsilon_F\pm i \eta)^2} \begin{pmatrix}
   -( \varepsilon_F\pm i\eta) & (\alpha+i\beta)(k_y+ik_x) \\
(\alpha-i\beta)(k_y-ik_x)& -( \varepsilon_F\pm i\eta)
\end{pmatrix},
\end{align}
\\where $\varepsilon_\textrm{F}$ is Fermi energy and the broadening parameter $\eta$ represents the inverse of the relaxation time $\tau$. Finally, the $ \langle S_x\rangle$ component in Eq.~\ref{A3} become
\begin{align}
\nonumber
    \langle S_x\rangle &= \frac{e E_y}{8\pi^3}\int_0^{\varepsilon_F/\sqrt{\alpha^2+\beta^2}}k dk\int_0^{2\pi} d\theta \frac{k^2 (\alpha^2+\beta^2)(\alpha \cos{2\theta}+\beta \sin{2\theta})-\alpha (\varepsilon_F^2+\eta^2)}{\varepsilon_F^4+2\varepsilon_F^2(\eta^2-(\alpha^2+\beta^2)k^2)+(k^2(\alpha^2+\beta^2)+\eta^2)^2}\\&= -\frac{e E_y \alpha (\varepsilon_F^2+\eta^2)}{16\pi^2\varepsilon_F (\alpha^2+\beta^2)\eta}\times (\tan^{-1}[\frac{\eta^2-\varepsilon_F^2+(\alpha^2+\beta^2)k^2}{2\varepsilon_F\eta}])\bigg|_{0}^{(\varepsilon_F/\sqrt{\alpha^2+\beta^2})} .
\end{align}

Similarly, the $y$ and $z$ components of spin polarisation (using Eq.~\ref{eq3} of the main text) are found to be

\begin{align}
\nonumber
\langle S_y\rangle &= \frac{e E_y}{8\pi^3}\int_0^{\varepsilon_F/\sqrt{\alpha^2+\beta^2}}k dk\int_0^{2\pi} d\theta \frac{k^2 (\alpha^2+\beta^2)(-\beta \cos{2\theta}+\alpha \sin{2\theta})+\beta (\varepsilon_F^2+\eta^2)}{\varepsilon_F^4+2\varepsilon_F^2(\eta^2-(\alpha^2+\beta^2)k^2))+(k^2(\alpha^2+\beta^2)+\eta^2)^2}\\&= \frac{e E_y \beta (\varepsilon_F^2+\eta^2)}{16\pi^2\varepsilon_F (\alpha^2+\beta^2)\eta}(\tan^{-1}[\frac{\eta^2-\varepsilon_F^2+(\alpha^2+\beta^2)k^2}{2\varepsilon_F\eta}])\bigg|_{0}^{(\varepsilon_F/\sqrt{\alpha^2+\beta^2})},
\end{align}

\begin{align}
     \langle S_z\rangle &= \frac{e E_y}{8\pi^3}\int_0^{\varepsilon_F/\sqrt{\alpha^2+\beta^2}}k dk\int_0^{2\pi} d\theta \frac{k (\alpha^2+\beta^2) \eta \cos{\theta}}{\varepsilon_F^4+2\varepsilon_F^2(\eta^2-(\alpha^2+\beta^2)k^2))+(k^2(\alpha^2+\beta^2)+\eta^2)^2}.
\end{align}
For $\eta \ll 2 \varepsilon_\textrm{F}$, the simplified form can be expressed as,
\begin{align}
\nonumber
    \langle S_x\rangle&\approx-\frac{e E_y \alpha \varepsilon_F}{32\pi(\alpha^2+\beta^2)\eta},\\
    \nonumber
    \langle S_y\rangle&\approx \frac{e E_y \beta \varepsilon_F}{32\pi (\alpha^2+\beta^2)\eta},\\\langle S_z\rangle&=0.
\end{align}
Thus, we derive the Eq.~(\ref{eq6}) of the main text, and the difference in the sign between $x$ and $y$ components is evident.  Next, we discuss the effect of the different relative strengths of RSOC and DSOC(+) on spin polarisation. 
\subsubsection{Case: $\alpha=\beta$}
When the strengths of RSOC and DSOC(+) are equal, the components of the spin polarisation become,
\begin{align}
\nonumber
    \langle S_x\rangle &\approx -\frac{e E_y \varepsilon_F}{64\pi \alpha \eta};~~
   \langle S_y\rangle\approx \frac{e E_y \varepsilon_F}{64\pi \alpha \eta};~~
    \langle S_z\rangle= 0.
\end{align}
\\
\subsubsection{Case: $\alpha \neq 0, \beta = 0$}
For only RSOC to be present in the system, the components of spin polarisation are obtained to be,
\begin{align}
     \langle S_x\rangle &\approx -\frac{e E_y \varepsilon_F}{32\pi \alpha \eta};~~
    \langle S_y\rangle =0 ;~~
   \langle S_z\rangle= 0.
\end{align}
This corroborates standard REE effect. 
\subsubsection{Case: $\alpha = 0, \beta \neq 0$}
In the presence of only DSOC(+), the components of spin polarisation can be written as,
\begin{align}
      \langle S_x\rangle &=0;~~
    \langle S_y\rangle\approx \frac{e E_y \varepsilon_F}{32\pi \beta \eta};~~
    \langle S_z\rangle= 0.\label{onlyDSOC+}
\end{align}
\\
\subsection{Electric field applied along x direction}
As discussed in the main text, the sign of the spin components of the spin polarisation depends on the direction of the applied field. To see, we now apply field along $x$, i. e., $\mathbf{E} = (E_x, 0, 0)$. With this, the components of the spin polarisation are given by

\begin{align}
\nonumber
    \langle S_x\rangle &= \frac{e E_x}{8\pi^3}\int_0^{\varepsilon_F/\sqrt{\alpha^2+\beta^2}}k dk\int_0^{2\pi} d\theta \frac{k^2 (\alpha^2+\beta^2)(\beta \cos{2\theta}-\alpha \sin{2\theta})+\beta (\varepsilon_F^2+\eta^2)}{\varepsilon_F^4+2\varepsilon_F^2(\eta^2-(\alpha^2+\beta^2)k^2)+(k^2(\alpha^2+\beta^2)+\eta^2)^2}\\
    &= \frac{e E_x \beta (\varepsilon_F^2+\eta^2)}{16\pi^2\varepsilon_F\eta (\alpha^2+\beta^2)}\times (\tan^{-1}[\frac{\eta^2-\varepsilon_F^2+(\alpha^2+\beta^2)k^2}{2\eta \varepsilon_F}])\bigg|_{0}^{(\varepsilon_F/\sqrt{\alpha^2+\beta^2})},
\end{align}
\begin{align}
\nonumber
    \langle S_y\rangle &= \frac{e E_x}{8\pi^3}\int_0^{\varepsilon_F/\sqrt{\alpha^2+\beta^2}}k dk\int_0^{2\pi} d\theta \frac{k^2 (\alpha^2+\beta^2)(\alpha \cos{2\theta}+\beta \sin{2\theta})+\alpha(\varepsilon_F^2+\eta^2)}{\varepsilon_F^4+2\varepsilon_F^2(\eta^2-(\alpha^2+\beta^2)k^2)+(k^2(\alpha^2+\beta^2)+\eta^2)^2}\\
    &= \frac{e E_x \alpha (\varepsilon_F^2+\eta^2)}{16\pi^2\varepsilon_F\eta (\alpha^2+\beta^2)}\times (\tan^{-1}[\frac{\eta^2-\varepsilon_F^2+(\alpha^2+\beta^2)k^2}{2\eta \varepsilon_F}])\bigg|_{0}^{(\varepsilon_F/\sqrt{\alpha^2+\beta^2})},
\end{align}
\begin{align}
\nonumber
     \langle S_z\rangle &= -\frac{e E_x}{8\pi^3}\int_0^{\varepsilon_F/\sqrt{\alpha^2+\beta^2}}k dk\int_0^{2\pi} d\theta \frac{k\eta(\alpha^2+\beta^2)  \sin{\theta}}{\varepsilon_F^4+2\varepsilon_F^2(\eta^2-(\alpha^2+\beta^2)k^2))+(k^2(\alpha^2+\beta^2)+\eta^2)^2}\\&= 0.
\end{align}
And for $\eta \ll 2 \varepsilon_\textrm{F}$, the components can be simplified as,
\begin{align}
\nonumber
    \langle S_x\rangle&\approx \frac{e E_x \beta \varepsilon_F}{32\pi\eta (\alpha^2+\beta^2)},\\
    \nonumber
    \langle S_y\rangle&\approx\frac{e E_x \alpha \varepsilon_F}{32\pi\eta(\alpha^2+\beta^2)},\\
    \langle S_z\rangle&=0.
\end{align}
Thus recover the Eq.~(\ref{eq7}) in the main text. Evidently, both the spin components have the same sign as opposed to the case discussed in the preceding section. For different respective strengths of RSOC and DSOC(+), the spin polarisation can be obtained very easily as before. 
\subsection{2DEG with only Dresselhaus terms}
We now compute the spin polarisation components for systems that feature only BIA, or in other words, systems with only linear DSOC terms in a 2DEG. We particularly focus on the DSOC(-) term and then compare it with the DSOC(+) term discussed in the preceding paragraph. As before, neglecting the kinetic energy term, the Hamiltonian for DSOC(-) is  
\begin{align}
H_{-} = \beta (k_x \sigma_x - k_y \sigma_y).
\end{align}
Then we find 
\begin{align}
\nonumber
    \langle S_y\rangle_{-} &= -\frac{e E_y \varepsilon_F}{32\pi \beta \eta},\\
   \langle S_x\rangle_{-}&= \langle S_z\rangle_{-}=0,
   \label{onlyDSOC-}
\end{align}
for the field along the $y$ direction. Here $``-"$ refers to contribution from the DSOC(-). Clearly, we see longitudinal spin polarisation with respect to the applied field similar to the DSOC(+) case (cf. Eq.~(\ref{onlyDSOC+})) but with an opposite sign. 

\section*{References:}
\bibliography{references}

\end{document}